\documentclass[11pt]{article}
\usepackage{amsmath,amsfonts,amssymb,amsthm}
\usepackage{graphicx}
\usepackage{braket}
\usepackage{geometry}
\usepackage{hyperref}
\title{When Recall Fails, Discord Remembers: A Quantum Analogue of Kuhn’s Theorem}
\author{Faisal Shah Khan\thanks{faisal\_khan@kenan-flagler.unc.edu.}}
\date{\today}

\begin{document}

\maketitle

\begin{abstract}
A behavioral quantum strategy is shown to replicate the payoff of a classical mixed strategy in an extensive-form game with imperfect recall, using only local measurements on a separable quantum state with zero entanglement and nonzero discord. Classical behavioral strategies, constrained by imperfect recall, cannot achieve this coordination. The result suggests a quantum analogue to Kuhn’s classical equivalence: discord enables behavioral-style strategies to functionally substitute for strategic memory and recover coordination lost in the classical setting. This highlights quantum discord as a minimal and robust resource for extending bounded rationality beyond classical limits.
\end{abstract}

\section*{Introduction}

Kuhn’s Theorem \cite{Kuhn} is a foundational result in classical game theory, demonstrating that in extensive-form games with perfect recall—where players remember all past choices and the information available at each point—mixed and behavioral strategies are outcome-equivalent \cite{Osborne}. Mixed strategies involve committing in advance to a probability distribution over complete plans of action. In contrast, behavioral strategies specify independent random choices at each decision node, allowing players to act locally without preplanning. Given perfect recall, the two approaches yield the same strategic outcomes: long-term planning can be replaced by appropriate on-the-spot decisions. This structure aligns with the concept of bounded rationality, as introduced by Herbert Simon \cite{Simon}, which emphasizes that real-world agents often operate under cognitive constraints, including limited memory and foresight. Behavioral strategies also simplify computation and can serve as psychological tools, introducing unpredictability into a player’s actions. 

\begin{figure}
    \centering
    \includegraphics[scale=0.75]{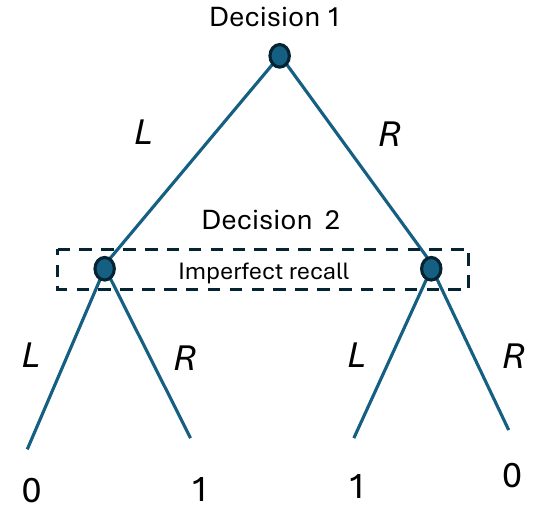}
    \caption{An imperfect recall game with two stages where the agent takes actions of $L$ versus $R$. The imperfect recall is depicted by the dashed box at decision node 2 where the player is unable to recall if they arrived at the one on the left or the one on the right. }
    \label{fig:extengame}
\end{figure}
In games with imperfect recall, however, Kuhn’s Theorem no longer holds. Behavioral strategies may underperform relative to mixed strategies, revealing the limits of classical models of bounded rationality when recall is compromised. This breakdown invites consideration of alternative mechanisms for coordinating decisions across stages without relying on memory or global planning. Quantum information theory \cite{Wilde} offers one such mechanism. While entanglement is often cited as the source of quantum advantage in strategic contexts \cite{Brandenburger, Horodecki}, recent work has shown that even separable states can exhibit non-classical correlations through a phenomenon known as quantum discord \cite{Ollivier, Henderson}.

Quantum discord captures the disturbance induced by local measurement, independent of entanglement, and offers a more minimal form of quantum correlation. The behavioral quantum strategy constructed in this paper uses a separable, discordant quantum state and local measurements to achieve the same payoff as a classical mixed strategy in an imperfect recall game. This coordination is accomplished without strategic recall, entanglement, or communication. While not a formal extension of Kuhn’s Theorem, the result suggests a quantum analogue: discord enables a behavioral-style strategy to recover the coordination power of mixing in a context where classical behavioral strategies fail.

Several authors have investigated the role of discord in strategic settings, including Nawaz, Wei, and Lowe \cite{Nawaz, Wei, Lowe}, though primarily in the context of normal-form games. This paper shifts the focus to extensive-form games with imperfect recall. The formalism of quantum extensive-form games introduced by Ikeda \cite{Ikeda} provides a structural basis for incorporating quantum transitions and interference into sequential decision-making, though it does not address discord. The present analysis demonstrates that quantum discord—unlike entanglement—can serve as a standalone operational resource for extending behavioral strategies beyond the limits imposed by classical information structures.

\section*{Classical and Quantum Strategies}

Consider a single-agent, extensive-form game with two sequential decisions of Figure \ref{fig:extengame}. At the first stage, the agent selects $a_1 \in \{L, R\}$. At the second stage, the agent selects $a_2 \in \{L, R\}$, but without recall of the first choice. Imperfect recall implies that the decision histories following $L$ or $R$ are indistinguishable at the second stage. The payoff function rewards alternating actions:
\begin{equation}
u(a_1, a_2) =
\begin{cases}
1 & \text{if } a_1 \neq a_2, \\
0 & \text{if } a_1 = a_2.
\end{cases}
\end{equation}

A behavioral strategy selects $a_1$ and $a_2$ independently. The best such strategy yields
\begin{equation}
\mathbb{E}[u] = \Pr(a_1 \neq a_2) = 0.5.
\end{equation}
In contrast, a mixed strategy can precommit to $(L,R)$ and $(R,L)$ with equal probability and produce a higher yield: 
\begin{equation}
\sigma = 0.5 \cdot (L,R) + 0.5 \cdot (R,L), \quad \mathbb{E}[u] = 1.
\end{equation}
A \emph{behavioral quantum strategy} achieving this optimal outcome without memory or communication uses the separable but discordant two-qubit quantum state:
\begin{equation}
\rho_{AB} = \frac{1}{2} \left( |0\rangle_A \langle 0| \otimes |+\rangle_B \langle +| \right)
+ \frac{1}{2} \left( |1\rangle_A \langle 1| \otimes |-\rangle_B \langle -| \right),
\end{equation}
with 
\begin{equation}
|+\rangle = \frac{1}{\sqrt{2}}(|0\rangle + |1\rangle), \quad
|-\rangle = \frac{1}{\sqrt{2}}(|0\rangle - |1\rangle)
\end{equation}
as the $X$-basis states. This state can be interpreted as the result of the agent locally preparing a separable but discordant correlation structure, either by deterministically or probabilistically applying appropriate local unitary transformations to the qubits (i.e. "flipping" them). The agent implements the strategy by performing local measurements independently on each qubit:

\begin{itemize}
\item Qubit $A$: measured in the computational basis $\{|0\rangle, |1\rangle\}$:
\[
|0\rangle \rightarrow a_1 = L, \quad |1\rangle \rightarrow a_1 = R.
\]

\item Qubit $B$: measured in the $X$-basis $\{|+\rangle, |-\rangle\}$:
\[
|-\rangle \rightarrow a_2 = L, \quad |+\rangle \rightarrow a_2 = R.
\]
\end{itemize}

Measurement on qubit $A$ collapses the system into either $|0\rangle$ or $|1\rangle$, which in turn projects qubit $B$ into $|+\rangle$ or $|-\rangle$. This procedure produces the outcomes $(L,R)$ and $(R,L)$ each with probability $0.5$, resulting in alternating actions in both cases. The expected payoff is therefore:
\begin{equation}
\mathbb{E}[u] = \frac{1}{2} \cdot u(L,R) + \frac{1}{2} \cdot u(R,L) = 1.
\end{equation}

Although the agent’s measurements are local and memoryless, the resulting outcomes are correlated due to quantum discord in the constructed state. While structurally similar to a classical behavioral strategy in being locally implemented, this approach violates the classical assumption of independence across decision points. These quantum correlations—arising without entanglement—enable outcome-level coordination in the absence of memory, planning, or communication. 

\section*{Quantum Discord in the Constructed Strategy}

The state \( \rho_{AB} \) used in the agent's behavioral quantum strategy 
\[
\rho_{AB} = \frac{1}{2} \, |0\rangle_A\langle 0| \otimes |+\rangle_B\langle +| + \frac{1}{2} \, |1\rangle_A\langle 1| \otimes |-\rangle_B\langle -|,
\]
and can be written as a convex combination of product states:
\begin{equation}
\rho_{AB} = \sum_{i=1}^{2} p_i \, \rho_A^i \otimes \rho_B^i,
\end{equation}
where \( p_1 = p_2 = \frac{1}{2} \), \( \rho_A^1 = |0\rangle_A\langle 0| \), \( \rho_B^1 = |+\rangle_B\langle +| \), \( \rho_A^2 = |1\rangle_A\langle 1| \), and \( \rho_B^2 = |-\rangle_B\langle -| \). This confirms that \( \rho_{AB} \) is separable and therefore has zero entanglement.

Despite the absence of entanglement, \( \rho_{AB} \) exhibits non-classical correlations, as captured by quantum mutual information. The reduced states are maximally mixed:
\[
\rho_A = \rho_B = \frac{1}{2} I \quad \Rightarrow \quad S(\rho_A) = S(\rho_B) = 1,
\]
where $S(.)$ is the von Neumann entropy, and since \( \rho_{AB} \) is a mixture of two orthogonal pure states, \( S(\rho_{AB}) = 1 \). Thus, the mutual information is
\begin{equation}
I(A : B) = S(\rho_A) + S(\rho_B) - S(\rho_{AB}) = 1 + 1 - 1 = 1.
\end{equation}

In contrast, a classical behavioral strategy implemented by independently tossing two fair coins produces a product distribution with zero mutual information: \( I_{behavioral}(A : B) = 0 \). The presence of mutual information in \( \rho_{AB} \) therefore signals correlations that cannot arise in the classical setting without memory or planning.

These correlations are further characterized by quantum discord, which distinguishes classical from genuinely quantum correlations in separable states. Discord is defined as the difference between mutual information and the classical correlation obtained after optimal local measurement. It is generally asymmetric, and its value depends on which subsystem is measured.

When a projective measurement is performed on qubit \( A \) in the computational basis \( \{|0\rangle_A, |1\rangle_A\} \), the post-measurement state of \( B \) becomes either \( |+\rangle_B \) or \( |-\rangle_B \), both pure states. The conditional entropy is thus zero, and the classical correlation measure is given by
\[
J(B|A) = S(\rho_B) - \sum_i p_i S(\rho_{B|i}) = 1 - 0 = 1,
\]
and therefore, the quantum discord is
\[
D(B|A) = I(A : B) - J(B|A) = 1 - 1 = 0.
\]

However, if qubit \( B \) is measured in the computational basis \( \{|0\rangle_B, |1\rangle_B\} \), it disturbs the superposition states \( |+\rangle_B \) and \( |-\rangle_B \), leading to mixed conditional states for \( A \). The resulting conditional entropy is nonzero, implying
\[
J(A|B) < 1 \quad \Rightarrow \quad D(A|B) = I(A : B) - J(A|B) > 0.
\]

This correlation, where the outcome of measuring qubit $A$ fully determines the post-measurement state of qubit $B$, mirrors the kind of dependence seen in entangled states, where measurement on one qubit instantaneously reveals the state of the other. However, in this case, the effect arises from quantum discord, not entanglement.

This confirms that \( \rho_{AB} \) possesses quantum discord: correlations that persist in separable states and resist classical explanation. In the agent’s strategy, this discord becomes operationally relevant under imperfect recall. After measuring qubit \( A \), the agent does not retain memory of the outcome, yet the subsequent measurement of qubit \( B \) yields coordinated actions. The payoff-optimal behavior, unattainable via classical memoryless strategies, is achieved through local, memoryless measurements on a discordant state—demonstrating how discord substitutes for recall and enables effective coordination without entanglement or communication. In addition to its operational utility under imperfect recall, quantum discord is also more robust to environmental noise than entanglement~\cite{Werlang}, making it a natural candidate for memory-free coordination in realistic settings.

\section*{Disscusion}

Quantum discord can be operationalized as a strategic resource in extensive-form games with imperfect recall. By constructing a separable, discordant state and implementing local measurements without memory or communication, an agent can replicate the outcomes of a classical mixed strategy that would otherwise be unreachable using behavioral strategies alone. This challenges the classical boundary between bounded rationality and full-information coordination, showing that non-classical correlations can substitute for memory in decision-making processes. In this sense, the result can be viewed as a quantum analogue of Kuhn’s Theorem: quantum discord enables a behavioral-style strategy to replicate the coordination power of a mixed strategy, even in an imperfect recall setting where classical behavioral strategies fail.

This strategic effect can be illustrated by imagining a robot navigating a two-stage decision task—such as a branching maze—where it must make two choices in sequence. Due to design constraints, the robot cannot recall its first decision when making the second. A classical robot in this setting would be limited to independent behavioral strategies, unable to coordinate its actions across stages. But if the robot is initialized with two parts of a separable, discordant quantum state, it can measure one part at the first node and the other at the second. Although no memory or communication is involved, the correlations induced by discord ensure that the second decision is appropriately conditioned on the first—functionally substituting for recall. The quantum structure provides coordination where classical bounded rationality would fail.

The behavioral quantum strategy constructed in this paper underscores a broader conceptual point: the strategic role of discord in this setting reflects a deeper link between non-classical correlation and informational uncertainty. 
Even in the absence of entanglement, separable states can limit an agent’s ability to extract information without inducing disturbance. This reveals discord as a signature of the irreducible informational asymmetries inherent in quantum systems. This nonclassical structure is operational: the advantage achieved by the behavioral quantum strategy using $\rho_{AB}$ cannot be replicated by any classical system without violating the assumption of imperfect recall. Although the state is separable, its discordant correlations enable perfect outcome-level coordination through local, memoryless measurements\footnote{While the coordination between local measurements in the quantum strategy may resemble nonlocal effects, the underlying quantum state is separable and does not violate Bell inequalities. The correlations are pre-established and entirely consistent with locality—no information is transmitted between measurements, and the construction does not enable signaling. The quantum advantage instead arises from how local measurements on a discordant state can induce conditionally pure post-measurement states. In this construction, measuring qubit \( A \) collapses \( B \) into \( |+\rangle \) or \( |-\rangle \), depending on the outcome—even though the global state is separable. Classical behavioral strategies under imperfect recall require statistically independent choices across stages. In contrast, the discordant structure allows local measurements to produce outcome-dependent coordination without recall or communication, effectively bypassing the classical independence constraint.}---something classically achievable only with recall or precommitment. This perspective echoes proposals in quantum economics, such as Qadir’s suggestion that the uncertainty principle may offer a more faithful model of economic and strategic behavior than classical probability alone \cite{Qadir}.

\end{document}